\documentclass[prd,showpacs,nofootinbib,preprintnumbers]{revtex4}

\usepackage{graphicx}
\usepackage{amssymb}
\usepackage{pifont}

\def\bfnabla{\mbox{\boldmath $\nabla$}}
\def\bfsigma{\mbox{\boldmath $\sigma$}}
\def\lQ{\Lambda_{\rm QCD}}
\newcommand{\be}{\begin{equation}}{\bf }
\newcommand{\ee}{\end{equation}}
\newcommand{\bea}{\begin{eqnarray}}
\newcommand{\eea}{\end{eqnarray}}
\def\als{\alpha_{\rm s}}
\def\siml{{\ \lower-1.2pt\vbox{\hbox{\rlap{$<$}\lower6pt\vbox{\hbox{$\sim$}}}}\ }}

\begin{document}

\title{Soft, collinear and non-relativistic modes in radiative decays of very heavy quarkonium}
\author{Xavier \surname{Garcia i Tormo}}
\author{Joan Soto}
\thanks{Member of CER  Astrophysics, Particle Physics and Cosmology, associated with Institut de Ci\`encies de l'Espai-CSIC.}
\affiliation{Departament d'Estructura i Constituents de la Mat\`eria, Universitat de Barcelona\\Diagonal 647, E-08028 Barcelona, Catalonia, Spain}
\preprint{UB-ECM-PF 03/39}



\begin{abstract}
We analyze the end-point region
 of the photon spectrum in semi-inclusive radiative decays of very
 heavy quarkonium ($m\als^2 \gg \Lambda_{QCD}$). We discuss the interplay of the scales arising in the
Soft-Collinear Effective Theory, $m$, $m(1-z)^{1/2}$ and $m(1-z)$ for $z$ close to $1$, with the scales of heavy
quarkonium systems in the weak coupling regime, $m$, $m\als$ and $m\als^2$. For $1-z \sim \als^2$ only collinear
 and (ultra)soft modes are seen to be relevant, but the recently discovered soft-collinear
 modes show up for $1-z \ll \als^2$.
The $S$- and $P$-wave octet shape functions are calculated. When they are included in the analysis of the photon spectrum of the $\Upsilon (1S)$ system, the agreement with data in the end-point region becomes excellent.
The NRQCD matrix elements $\left<\phantom{}1^3S_1\vert
O_8(\phantom{}^1S_0) \vert \phantom{}1^3S_1 \right>$ and
$\left<\phantom{}1^3S_1\vert
O_8(\phantom{}^3P_J) \vert \phantom{}1^3S_1 \right>$ are also obtained.

\end{abstract}


\maketitle

\section{Introduction}

Effective field theories (EFT) have proved extremely useful in the field of strong interactions. Applications to
high energy processes in QCD involving  very energetic partons \cite{duncan}, however, have been elusive until
recently. Important features of a suitable EFT for such processes were outlined in \cite{luke}, which led to the
development of the so called
 Soft-Collinear Effective Theory (SCET) \cite{bauer,beneke,neubert} (see \cite{school} for a pedagogical introduction).

SCET has generated high expectations. Indeed factorization proofs appear to greatly simplify and power corrections seem to become under control.
In addition,
a large number of potential applications is envisaged \cite{ira}. Among those, exclusive and semi-inclusive $B$-decays have deserved special attention due to the necessity to have a
 good control on the hadronic effects in order to extract the
CKM matrix elements from the abundant
 $B$-factory data.

SCET was originally formulated in terms of soft, collinear and usoft modes. Later, it was realized that two possible scalings for collinear modes were relevant and the terminology SCET$_{\rm I}$ and SCET$_{\rm II}$ was introduced. Recently a new mode,
called soft-collinear, has been claimed to be necessary \cite{neubert} (see \cite{new} for the latest discussions). It is often assumed that some of the momentum components of these modes
have typical sizes $\sim \lQ$ or even smaller.

One of the difficulties that one faces in $B$-physics is that the
bound state dynamics of
the initial $B$-meson
is
dominated by the scale $ \Lambda_{\rm QCD}$, and hence a weak coupling analysis is not reliable. Therefore, the interplay of the initial bound state dynamics with final state modes of momentum components of the order $\lQ$ (or smaller) is difficult to
 figure out.
We advocate here that a very heavy quarkonium in the initial state may provide an excellent theoretical tool
 to shed light on this issue, since the bound state dynamics occurs at weak coupling and it is amenable of a detailed
 analysis.
We shall illustrate this point by analyzing the end-point region of the photon spectrum in inclusive decays of
very heavy quarkonium.

Semi-inclusive radiative decays for the $\Upsilon (1S)$
have already been discussed in the framework of SCET \cite{many,FL0,FL}.
 SCET has been used to put forward factorization formulas and to resum Sudakov logarithms. An improved description of data \cite{cleo} with respect to earlier approaches \cite{earlier} has been achieved.
However, the bound state dynamics, which is relevant for the evaluation of the octet shape functions, has not been
studied in detail, but rather modeled by analogy with $B$-meson systems \cite{neubertmodel}, which is a doubtful
approximation. We shall calculate here the octet shape functions under the assumption that the bottom quark is
sufficiently heavy as
 to consider $\Upsilon (1S)$ a Coulombic state. This assumption appears to be self-consistent in the calculation
 of the spectrum \cite{yndurain,sumino,hamburg,pinedahamburg}, and decay and production currents \cite{currenthamburg}. We observe
that the factorization scale dependence of the shape functions is sensible to the bound state dynamics and discuss
its cancellation.

We distribute the paper as follows. In section \ref{phespwcr} we calculate the photon spectrum at the end-point region in the weak coupling regime. We do so
 by matching first QCD to Non-Relativistic QCD (NRQCD) \cite{BBL} +SCET$_{\rm I}$, next NRQCD+SCET$_{\rm I}$ to Potential NRQCD (pNRQCD) \cite{mont}+SCET$_{\rm II}$, and
finally carrying out the calculations in the later EFT. We confirm the factorization formulas \cite{FL0,FL} and obtain the
octet shape functions.
In section \ref{appl}, we apply our results to the $\Upsilon (1S)$ system and obtain a very good description of the
 experimental data \cite{cleo} in the end-point region.
 In section \ref{disc} we discuss the interplay of the several scales in the problem, in particular the emergence of a soft-collinear mode.
Section \ref{concl} is devoted to the conclusions.
In the Appendix
we present results for NRQCD octet matrix elements of the $1\phantom{}^3S_1$ state, which follow
from those of section \ref{phespwcr}.

\section{The end-point region of the photon spectrum
}\label{phespwcr}

We start from the formula given in Ref. \cite{FL}

\begin{equation}\label{G}
{d \Gamma\over dz}=z{M\over 16\pi^2} {\rm Im} T(z)\quad \quad T(z)=-i\int d^4 x e^{-iq\cdot x}\left< \phantom{}n^{2S+1}L_{J} \vert T\{ J_{\mu} (x) J_{\nu} (0)\} \vert \phantom{}n^{2S+1}L_{J} \right> \eta^{\mu\nu}_{\perp}
\end{equation}
where $J_{\mu} (x)$ is the electromagnetic current for heavy quarks in QCD and we have used spectroscopic notation for the heavy quarkonium states. The formula above holds for states fulfilling relativistic normalization. In the case that non-relativistic normalization is used, as we shall do below, the rhs of either the first or second formulas in (1) must be multiplied by $2M$, $M$ being the mass of the heavy quarkonium state. At the end-point region
the photon momentum (in light cone coordinates) in the rest frame of the heavy quarkonium is $q=\left(q_{+},q_{-}, q_{\perp}\right)=(zM/2,0,0)$ with $z \sim 1$ ($M\sqrt{1-z} \ll M$). This together with
 the fact that the heavy quarkonium is a non-relativistic system fixes the relevant kinematic situation. It is precisely in this situation when the standard NRQCD factorization (operator product expansion) breaks down \cite{rotwise}. The quark (antiquark)
 momentum
in the $Q\bar Q$ rest frame can be written as $p=(p_0, {\bf p})$, $p_0=m+l_0, {\bf p}={\bf l}$;
 $l_0 , {\bf l} \ll m$, $m$ being the mass of the heavy quark ($M\sim 2m$).
Momentum conservation implies that if a few gluons are produced in the short distance annihilation process at
least one of them has momentum $r=( r_{+},r_{-}, r_{\perp})$, $r_{-} \sim M/2$ ; $r_{+}, r_{\perp}\ll M$,
 which we will call collinear. At short distances, the emission of hard gluons is
penalized by $\als (m)$ and the
 emission of softer ones by powers of  soft scale over $M $. Hence, the leading contribution at short
distances
 consists of the emission of a single collinear gluon. This implies that the   $Q\bar Q$ pair must be in a color octet configuration, which means that the full process will have an extra long distance suppression related to
 the emission
of (ultra)soft gluons. The next-to-leading contribution at short distances already allows for a singlet $Q\bar Q$
configuration. Hence, the relative weight of color-singlet and color-octet configurations depends
not only on $z$ but also on the
bound state dynamics, and it is difficult to establish a priori. In order to do so, it is advisable to implement the
constraints above by introducing suitable  EFTs. In a first stage we need NRQCD \cite{BBL}, which factors out the
scale $m$ in the $Q\bar Q$ system, supplemented with collinear gluons,
namely gluons for which
the scale $m$ has been factored out
 from the components
$r_{+}, r_{\perp} $ (but is still active in the component $r_{-}$). For the purposes of this work it is enough to take for the Lagrangian of the collinear gluons the full QCD Lagrangian and enforce  $r_{+}, r_{\perp} \ll m$ when necessary.

\subsection{Matching QCD to NRQCD+SCET$_{\rm I}$}\label{NS}

For definiteness, we shall restrict our analysis to $^3 S_1$ states,
which decay mainly through two additional gluons.
At tree level, the electromagnetic current in (\ref{G}) can be matched to
the following currents in this EFT \cite{FL}\footnote{One loop matching calculations are already available analytical for the octet currents \cite{malpetr} and numerical for the singlet one \cite{kramer}.}

\begin{equation}\label{efcur}
J_{\mu} (x)= e^{-i2mx_0}\left( \Gamma_{\alpha\beta i\mu}^{(1,\phantom{}^3S_1)}J^{i\alpha\beta}_{(1,\phantom{}^3S_1)} (x)
+\Gamma_{\alpha\mu}^{(8,\phantom{}^1S_0)}J^{\alpha}_{(8,\phantom{}^1S_0)} (x)+
\Gamma_{\alpha\mu ij}^{(8,\phantom{}^3P_J)}J^{\alpha ij}_{(8,\phantom{}^3P_J)} (x) + \dots
\right) + h.c.
\end{equation}

\bea
 & \Gamma_{\alpha\beta i\mu}^{(1,\phantom{}^3S_1)}={g_s^2 e e_Q\over 3 m^2}\eta^{\perp}_{\alpha\beta}\eta_{\mu i} &J^{i\alpha\beta}_{(1,\phantom{}^3S_1)} (x)= \chi^{\dagger}\bfsigma^{i} \psi Tr\{ B^{\alpha}_{\perp}  B^{\beta}_{\perp}\}(x) \cr & & \cr
&\Gamma_{\alpha\mu}^{(8,\phantom{}^1S_0)}= {g_s e e_Q\over m} \epsilon_{\alpha\mu}^{\perp}&J^{\alpha}_{(8,\phantom{}^1S_0)} (x)=  \chi^{\dagger}B^{\alpha}_{\perp} \psi (x)\cr & & \cr
&\Gamma_{\alpha\mu ij}^{(8,\phantom{}^3P_J)}= {g_s e e_Q\over m^2}\left(\eta_{\alpha j}^{\perp} \eta_{\mu i}^{\perp} +\eta_{\alpha i}^{\perp} \eta_{\mu j}^{\perp}-\eta_{\alpha\mu}^{\perp} n^{j} n^{i}\right) & J^{\alpha i j}_{(8,\phantom{}^3P_J)} (x)= -i \chi^{\dagger}B^{\alpha}_{\perp} \bfnabla^{i}\bfsigma^{j}\psi (x)
\eea
where $n=(n_+,n_-,n_{\perp})=(1,0,0)$ and $\epsilon_{\alpha\mu}^{\perp}=\epsilon_{\alpha\mu\rho 0}n^\rho $. These effective currents can be identified with the leading order in $\als$ of the currents introduced in \cite{FL}.
We use both Latin ($1$ to $3$) and Greek ($0$ to $3$) indices, $B^{\alpha}_{\perp}$ is a single collinear gluon field here, and $ee_Q$ is the charge of the heavy quark. Note,
however, that in order to arrive at (\ref{efcur}) one need not specify the scaling of collinear
fields as
$M(\lambda^2, 1, \lambda)$
but only the cut-offs mentioned
above, namely $r_{+}, r_{\perp}
\ll M$. Even though the $P$-wave octet piece appears to be $1/m$ suppressed with
respect to the $S$-wave octet piece, it will eventually give rise to contributions of the same order once the
bound state effects are taken into account. This is due to the fact that the $^3S_1$ initial state needs a chromomagnetic transition to become an octet $^1S_0$, which is $\als$ suppressed with respect to the chromoelectric transition required to become an octet $^3P_J$.

$T(z)$ can then be written as
\bea\label{T}
T(z)&=&H^{(1,\phantom{}^3S_1)}_{ii'\alpha\alpha'\beta\beta'}T_{(1,\phantom{}^3S_1)}^{ii'\alpha\alpha'\beta\beta'}+H^{(8,\phantom{}^1S_0)}_{\alpha\alpha^{\prime}}T_{(8,\phantom{}^1S_0)}^{\alpha\alpha^{\prime}}+
H^{(8,\phantom{}^3P_J)}_{\alpha ij\alpha^{\prime}i'j'}T_{(8,\phantom{}^3P_J)}^{\alpha ij\alpha^{\prime}i'j'}
+\cdots
\eea
where
\bea
H^{(1,\phantom{}^3S_1)}_{ii'\alpha\alpha'\beta\beta'}&=& \eta_{\perp}^{\mu\nu}\Gamma_{\alpha\beta i\mu}^{(1,\phantom{}^3S_1)}\Gamma_{\alpha'\beta'i'\nu}^{(1,\phantom{}^3S_1)}\nonumber\\
H^{(8,\phantom{}^1S_0)}_{\alpha\alpha^{\prime}}&=& \eta_{\perp}^{\mu\nu}\Gamma_{\alpha\mu}^{(8,\phantom{}^1S_0)}\Gamma_{\alpha'\nu}^{(8,\phantom{}^1S_0)}\nonumber\\
H^{(8,\phantom{}^3P_J)}_{\alpha ij\alpha^{\prime}i'j'}&=& \eta_{\perp}^{\mu\nu}\Gamma_{\alpha\mu ij}^{(8,\phantom{}^3P_J)}\Gamma_{\alpha'\nu i'j'}^{(8,\phantom{}^3P_J)}
\eea
and
\bea
T_{(1,\phantom{}^3 S_1)}^{ii'\alpha\alpha'\beta\beta'}(z)&= &-i\int d^4 x e^{-iq\cdot x-2mx_0}\left<\phantom{}^{3}S_{1} \vert T\{ {J^{i\alpha\beta}_{(\phantom{}1,^3 S_1)} (x)}^{\dagger} J^{i'\alpha^{\prime}\beta^{\prime}}_{(1,\phantom{}^3 S_1)} (0)\}
 \vert \phantom{}^{3}S_{1} \right> & \cr
 T_{(8,\phantom{}^1 S_0)}^{\alpha\alpha^{\prime}}(z)&= &-i\int d^4 x e^{-iq\cdot x-2mx_0}\left<\phantom{}^{3}S_{1} \vert T\{ {J^{\alpha}_{(8,^1 S_0)} (x)}^{\dagger} J^{\alpha^{\prime}}_{(8,\phantom{}^1 S_0)} (0)\}
 \vert \phantom{}^{3}S_{1} \right> & \cr
T_{(8,\phantom{}^3 P_J)}^{\alpha ij\alpha^{\prime}i'j'}(z)&= &-i\int d^4 x e^{-iq\cdot x-2mx_0}\left<\phantom{}^{3}S_{1} \vert T\{ {J^{\alpha ij}_{(8,\phantom{}^3 P_J)} (x)}^{\dagger} J^{\alpha^{\prime}i'j'}_{(8,\phantom{}^3 P_J)} (0)\}
 \vert \phantom{}^{3}S_{1} \right>
\eea
In (\ref{T}) we have not written a crossed term $(8,\phantom{}^1S_0$-$\phantom{}^3P_J)$ since it eventually vanishes at the order we will be calculating.

\subsection{Matching NRQCD+SCET$_{\rm I}$ to pNRQCD+SCET$_{\rm II}$}

If we restrict ourselves to $z$ such that $M(1-z)\siml m\als^2$, the scale of the binding energy, we can proceed
one step further in the EFT hierarchy. As discussed in Ref. \cite{mont}, NRQCD still contains quarks and gluons
with energies $\sim m\als$, which, in the situation above, can be integrated out. This leads to Potential NRQCD
(pNRQCD). On the SCET side, the restriction above implies that one may also restrict collinear gluons in the final
state to have $r_+ , r_{\perp} \ll M\sqrt{1-z} \siml m\als $, as we shall do. The scale $M\sqrt{1-z}$, which is
still active in SCET$_{\rm I}$, must then be integrated out \cite{many}. The integration of this scale produces the
dominant contributions from the color singlet currents. We have
\begin{displaymath}
\left<\phantom{}^{3}S_{1} \vert T\{ {J^{i\alpha\beta}_{(\phantom{}1,^3 S_1)} (x)}^{\dagger} J^{i'\alpha^{\prime}\beta^{\prime}}_{(1,\phantom{}^3 S_1)} (0)\}
 \vert \phantom{}^{3}S_{1} \right> \longrightarrow
\end{displaymath}
\begin{equation}
\longrightarrow 2N_c
{S_{V}^{i}}^{\dagger}({\bf x}, {\bf 0}, x_0)S_{V}^{i'}({\bf 0}, {\bf 0}, 0)
\left< \textrm{{\small VAC}}\vert Tr\{ B^{\alpha}_{\perp}  B^{\beta}_{\perp}\}(x)
Tr\{ B^{\alpha^{\prime}}_{\perp}  B^{\beta^{\prime}}_{\perp}\}(0)\vert \textrm{{\small VAC}}\right>\label{hardcoll}
\end{equation}
The calculation of the vacuum correlator for collinear gluons above has been carried out in \cite{FL}, and the final result, which is obtained by sandwiching (\ref{hardcoll}) between the quarkonium states, reduces to the one put forward in that reference.

For the color octet currents, the leading contribution arises from
a tree level
matching of the currents (\ref{efcur}),

\bea\label{efcur2}
J^{\alpha}_{(8,\phantom{}^1S_0)} (x) &\longrightarrow &\sqrt{2T_F} O_{P}^a ({\bf x}, {\bf 0}, x_0)B^{a \alpha}_{\perp}(x)\cr & & \cr
J^{\alpha ij}_{(8,\phantom{}^3P_J)} (x)& \longrightarrow &\sqrt{2T_F}\left.\left( i\bfnabla^i_{\bf y} O_{V}^{a j}({\bf x}, {\bf y}, x_0)\right)\right\vert_{{\bf y}={\bf 0}} B^{a \alpha}_{\perp}(x)
\eea
$S_{V}^{i}$, $O_{V}^{a i}$ and $O_{P}^a$ are the projection of the
singlet and octet wave function fields introduced in \cite{mont} to their vector and pseudoscalar components, namely $S=(S_{P}+S_{V}^{i}\sigma^i)/\sqrt{2}$
and $O^a =(O_{P}^{a}+O_{V}^{a i}\sigma^i)/\sqrt{2}$. $T_F=1/2$ and $N_c=3$ is the number of colors. $B^{a \alpha}_{\perp}(x)$ in (\ref{efcur2}) are now collinear gluons with $r_+ , r_{\perp} \ll M\sqrt{1-z} \siml m\als $.

\subsection{Calculation in pNRQCD+SCET$_{\rm II}$}\label{cpns}

\begin{figure}
\centering
\includegraphics{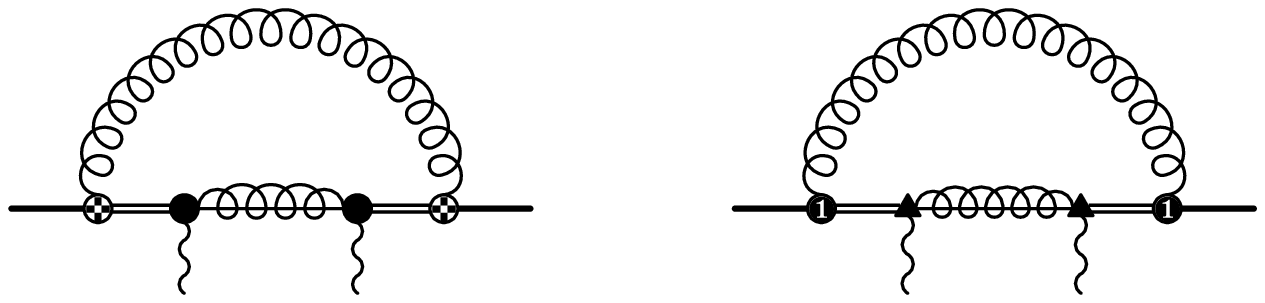}
\caption{\label{dos}Color octet contributions. $\bullet$ represents the color octet S-wave current,
$\blacktriangle$ represents the color octet P-wave current. The notation for the other vertices is
that of Ref. \cite{dib}, namely \ding{60}:= ${ig c_F \over \sqrt{N_c T_F}} { \left ( {\bfsigma}_1 - {\bfsigma}_2 \right ) \over 2m } \, {\rm Tr} \left [ T^b {\bf B} \right ]$ and \ding{182}:= ${ig\over \sqrt{N_c T_F}} {\bf x} \, {\rm Tr} \left [ T^b {\bf E} \right ]$. The solid line represents the singlet field, the double line represents the octet field and the gluon with a line inside represents a collinear gluon.}
\end{figure}

We shall then calculate the contributions of the color octet currents in pNRQCD coupled to collinear gluons. They
are depicted in Fig. \ref{dos}. For the contribution of the $P$-wave current, it is enough to have the pNRQCD Lagrangian at leading (non-trivial) order in the multipole expansion given in \cite{mont}. For the contribution of the $S$-wave current, one needs a $1/m$ chromomagnetic term given in \cite{dib}.

Let us consider the contribution of the $S$-wave color octet current in some detail. We have from the first diagram of Fig. \ref{dos},
\begin{displaymath}
 T_{(8,\phantom{}^1S_0)}^{\alpha\alpha^{\prime}}(z)=
-i\eta^{\alpha\alpha^{\prime}}_{\perp}(4\pi){32\over 3}T_F^2 \left({ c_F\over 2m}\right)^2 \als (\mu_u)C_f\int d^3 {\bf x}  \int d^3 {\bf x}^{\prime}
\psi^{\ast}_{n0}( {\bf x}^{\prime}) \psi_{n0}( {\bf x})\int\!\!\! {d^4 k\over (2\pi)^4} {{\bf k}^2\over k^2+i\epsilon}\times
\end{displaymath}
\begin{equation}\label{eqonas}
\times\left(1\over -k_0+E_n-h_o+i\epsilon\right)_{{\bf x}^{\prime},{\bf 0}}{1\over (M(1-z)-k_{+})M-{\bf k}_{\perp}^2+i\epsilon}
 \left(1\over -k_0+E_n-h_o+i\epsilon\right)_{{\bf 0}, {\bf x}}
\end{equation}
where we have used the Coulomb gauge (both for ultrasoft and collinear gluons). $E_n < 0$ is the binding energy ($M=2m+E_n$) of the heavy quarkonium, $\psi_{n0}( {\bf x})$ its wave function, and $h_o$ the color-octet Hamiltonian at leading order, which contains the kinetic term and a repulsive Coulomb potential \cite{mont}. $c_F$ is the hard matching coefficient of the chromomagnetic interaction in NRQCD \cite{BBL}, which will eventually be taken to $1$. We have also enforced that $k$ is ultrasoft by neglecting it in front of $M$ in the collinear gluon propagator.
We shall evaluate (\ref{eqonas}) in light cone coordinates.  If we carry out first the integration over $k_{-}$, only the pole
$k_{-}={\bf k}_{\perp}^2/k_{+}$ contributes. Then the only remaining singularities in the integrand are in the collinear gluon
 propagator. Hence, the absorptive piece can only come from its pole $M^2(1-z)-M k_{+}= {\bf k}_{\perp}^2$.
If $k_{+} \siml M(1-z)$, then  ${\bf k}_{\perp}^2 \sim M^2(1-z)$ which implies $k_{-}\sim M$. This contradicts the assumption that $k$ is ultrasoft. Hence,
${\bf k}_{\perp}^2$ must be expanded in the collinear gluon propagator. We then have
\begin{displaymath}
{\rm Im}\left(T_{(8,\phantom{}^1S_0)}^{\alpha\alpha^{\prime}}(z)\right)=-\eta^{\alpha\alpha^{\prime}}_{\perp}(4\pi){32\over 3}T_F^2 \left({ c_F\over 2m}\right)^2 \als(\mu_u)C_f\times
\end{displaymath}
\begin{displaymath}
\times\int d^3 {\bf x} \int d^3 {\bf x}^{\prime}
\psi^{\ast}_{n0}( {\bf x}^{\prime}) \psi_{n0}( {\bf x}){1\over 8\pi M}\int_0^{\infty} dk_+ \delta \left(M(1-z) -k_+\right)\times
\end{displaymath}
\begin{equation}\label{imts}
\times\int_0^{\infty} dx \left( \left\{ \delta ({\bf \hat x}) , {h_o-E_n \over h_o -E_n +{k_+ \over 2}+x} \right\} - {h_o-E_n \over h_o -E_n +{k_+ \over 2}+x}\delta ({\bf \hat x}){h_o-E_n \over h_o -E_n +{k_+ \over 2}+x} \right)_{{\bf x},{\bf x}^{\prime}}
\end{equation}
where we have introduced the change of variables $\vert {\bf k}_{\perp}\vert=\sqrt{2k_+x}$. Restricting ourselves to the ground state ($n=1$) and using the techniques of reference \cite{benekeschuler} we obtain
\begin{displaymath}
{\rm Im}\left(T_{(8,\phantom{}^1S_0)}^{\alpha\alpha^{\prime}}(z)\right)=-\eta^{\alpha\alpha^{\prime}}_{\perp}{16\over 3}T_F^2 \left({ c_F\over 2m}\right)^2 \als(\mu_u)C_f{1\over M}\int_0^{\infty} dk_+ \delta (M(1-z) -k_+)\times
\end{displaymath}
\begin{displaymath}
\times\int_0^{\infty} dx \left( 2 \psi_{10}( {\bf 0})I_{S}({k_+\over 2} +x)- I_{S}^2({k_+\over 2} +x) \right)
\end{displaymath}
\begin{displaymath}
I_{S}({k_+\over 2} +x):=\int d^3 {\bf x} \psi_{10}( {\bf x})\left( {h_o-E_1 \over h_o -E_1 +{k_+ \over 2}+x}\right)_{{\bf x},{\bf 0}}=
\end{displaymath}
\begin{equation}
=m\sqrt{\gamma\over \pi}{\als N_c \over 2}{1\over 1-z'}\left( 1-{2z'\over 1+z'} \;\phantom{}_2F_1\left(-\frac{\lambda}{z'},1,1-\frac{\lambda}{z'},\frac{1-z'}{1+z'}\right)\right)
\end{equation}
where
\begin{equation}
\gamma=\frac{mC_f\als}{2}\quad z'=\frac{\kappa}{\gamma}\quad-\frac{\kappa^2}{m}=E_1-\frac{k_+}{2}-x\quad\lambda=-\frac{1}{2N_cC_f}
\end{equation}
($E_1=-m\frac{\left(C_f\als\right)^2}{4}=-{\gamma^2\over m}$). This result can be recast in the factorized form given in \cite{FL}.
\begin{displaymath}
{\rm Im}\left(T_{(8,\phantom{}^1S_0)}^{\alpha \alpha^{\prime}}(z)\right)=-\eta^{\alpha \alpha^{\prime}}_{\perp}\int dl_+ S_{S}(l_+)
{\rm Im} J_M (l_+ - M(1-z))
\end{displaymath}
\begin{displaymath}
{\rm Im} J_M(l_+ - M(1-z))=
T_F^2\left(N_c^2-1\right){2\pi\over M}\delta (M(1-z) -l_+)
\end{displaymath}
\begin{equation}\label{resonas}
S_{S}(l_+)={4\als (\mu_u)\over 3 \pi N_c} \left({ c_F\over 2m}\right)^2
\int_0^{\infty} dx \left( 2 \psi_{10}( {\bf 0})I_{S}({l_+\over 2} +x)- I_{S}^2({l_+\over 2} +x) \right)
\end{equation}
We have thus obtained the $S$-wave color octet shape function $S_{S}(l_+)$.
Analogously, for the $P$-wave color octet shape functions, we obtain from the second diagram of Fig. \ref{dos}
\begin{displaymath}
{\rm Im}\left(T_{(8,\phantom{}^3P_J)}^{\alpha i j \alpha^{\prime} i' j'}(z)\right)\!\!\!=\!\!-\eta_{\perp}^{\alpha \alpha^{\prime}}\delta^{j j'}\!\!\!\int\!\!\! dl_+\!\! \left( \delta_{\perp}^{i i'}S_{P1}(l_+)+\left(n^i n^{i'}\!\!-{1\over 2}\delta_{\perp}^{i i'}\right)S_{P2}(l_+)\right)
{\rm Im} J_M(l_+ - M(1-z))
\end{displaymath}
\begin{displaymath}
S_{P1}(l_+):=
{\als (\mu_u)\over 6 \pi N_c}
\int_0^{\infty}\!\!\!dx\left( 2\psi_{10}( {\bf 0})I_P(\frac{l_+}{2}+x)-I_P^2(\frac{l_+}{2}+x) \right)
\end{displaymath}
\begin{equation}\label{ImTp}
S_{P2}(l_+):=
{\als (\mu_u)\over 6 \pi N_c}
\int_0^{\infty}\!\!\!dx \frac{8l_+x}{\left(l_++2x\right)^2}\left(
\psi^2_{10}( {\bf 0})-2\psi_{10}( {\bf 0})I_P(\frac{l_+}{2}+x)+I_P^2(\frac{l_+}{2}+x)\right)
\end{equation}
where
\begin{displaymath}
I_{P}({k_+\over 2} +x):=-\frac{1}{3}\int d^3 {\bf x} {\bf x}^i \psi_{10}( {\bf x})\left( {h_o-E_1 \over h_o -E_1 +{k_+ \over 2}+x} \bfnabla^i \right)_{{\bf x},{\bf 0}}=
\end{displaymath}
\begin{displaymath}
=\sqrt{\frac{\gamma^3}{\pi}}
{8\over 3}\left(
2-\lambda \right)\!\!\frac{1}{4(1+z')^3}\Bigg( 2(1+z')(2+z')+(5+3z')(-1+\lambda)+2(-1+\lambda)^2+
\end{displaymath}
\begin{equation}
\left.+\frac{1}{(1-z')^2}\left(4z'(1+z')(z'^2-\lambda^2)\left(\!\!-1+\frac{\lambda(1-z')}{(1+z')(z'-\lambda)}+\phantom{}_2F_1\left(-\frac{\lambda}{z'},1,1-\frac{\lambda}{z'},\frac{1-z'}{1+z'}\right)\right)\right)\right)
\end{equation}
Note that two shape functions are necessary for the $P$-wave case.
The three shape functions above are UV divergent and need regularization and renormalization. In order to regulate them at this order it is
enough to calculate the ultrasoft loop (the integral over $k$ in (\ref{eqonas})) in $D$-dimensions (leaving the bound state dynamics in $3$ space dimensions).
The UV behavior can be easily obtained by making an expansion of $I_S$ and $I_P$ in $1/z^{\prime}$, which is displayed in formulas (\ref{IStaylor}) and (\ref{IPtaylor}) of the Appendix. For the purpose of this section we only need the expansions up to order $1/{z^{\prime}}^2$.
The singular pieces read ($D=4-2\varepsilon$)
\begin{displaymath}
\left.S_{S}(l_+)\right\vert_{\varepsilon\rightarrow 0}\simeq
{4c_F^2\als (\mu_u )\gamma^5\over 3\pi^2 N_c m^3}(1-\lambda)\left( -2+\lambda (2\ln2 + 1)\right)\left(\frac{1}{\varepsilon}+\ln\left(\frac{\mu}{\frac{l_+}{2}+\frac{\gamma^2}{m}}\right)+\cdots\right)
\end{displaymath}
\begin{displaymath}
\left.S_{P1}(l_+)\right\vert_{\varepsilon\rightarrow 0}\simeq
{4\als (\mu_u )\gamma^5\over 9\pi^2 N_c m}(2-\lambda )
\left(-\frac{17}{6}+\lambda (2\ln2+\frac{1}{6})\right)\left(\frac{1}{\varepsilon}+\ln\left(\frac{\mu}{\frac{l_+}{2}+\frac{\gamma^2}{m}}\right)+\cdots\right)
\end{displaymath}
\begin{equation}
\left.
S_{P2}(l_+)\right\vert_{\varepsilon\rightarrow 0}\simeq
{\als (\mu_u ) l_+ \gamma^3\over 3\pi^2 N_c}
\label{singular}
\left(\frac{1}{\varepsilon}
+\ln\left(\frac{\mu}{l_+}\right)+\cdots
\right)
\end{equation}

\begin{figure}
\centering
\includegraphics{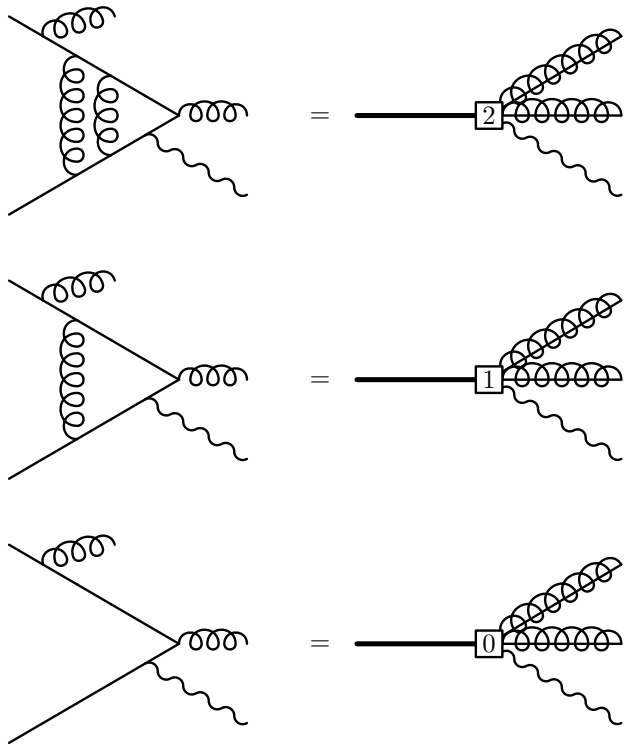}
\caption{\label{match}Relevant diagrams in the matching calculation QCD $\rightarrow$ pNRQCD+SCET.}
\end{figure}

\begin{figure}
\centering
\includegraphics{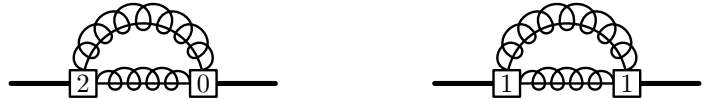}
\caption{\label{div}Diagrams which induce an IR scale dependence which cancels against the UV one of the octet shape functions.
}
\end{figure}

The renormalization is not straightforward. We will assume that suitable operators exists which may absorb the $1/\varepsilon$ poles so that a MS scheme makes sense to define the above expressions, and discuss in the following the origin of such operators. In order to understand the scale dependence of (\ref{singular}) it is  important to notice that it appears because the term
${\bf k}_{\perp}^2$ in the collinear gluon propagator is neglected in (\ref{eqonas}). It should then cancel with an IR divergence induced
by keeping the term ${\bf k}_{\perp}^2$, which implies assuming a size $M^2(1-z)$ for it, and expanding the ultrasoft
scales accordingly. We have checked that it does. However, this contribution cannot be computed reliably within
pNRQCD (neither within NRQCD) because it implies that the $k_{-}$ component of the ultrasoft gluon is of order $M$,
and hence it becomes collinear. A reliable calculation involves (at least) two steps within the EFT strategy. The first one
is the matching calculation of the singlet electromagnetic current at higher orders both in $\als$ and in $({\bf k}_{\perp}/M)^2$ and $k_+/M$. The second is a one loop calculation with collinear gluons involving the higher order singlet currents. Notice, before going on, that all divergences (and logarithms) of $S_S (l_+)$ and $S_{P1} (l_+)$ in (\ref{singular}) are sensible to the bound state dynamics, whereas those for $S_{P2} (l_+)$ are not. For the former, Fig. \ref{match} shows the relevant diagrams
which contribute to the IR behavior we are eventually looking for. We need NNLO in $\als$, but only LO in the $({\bf k}_{\perp}/M)^2$ and $k_+/M$ expansion.
These diagrams are IR finite, but they induce, in
 the second step, the
IR behavior which matches the UV of (\ref{singular}). The second step
amounts to integrating out the scale $M\sqrt{1-z}$ by
calculating  the loops with collinear gluons and expanding smaller scales in the integrand. We have displayed in
Fig. \ref{div} the two diagrams which provide the aforementioned IR divergences.
For the latter,
the UV behavior of which does not depend on the bound state dynamics, we need the matching at LO in $\als$ (last diagram in Fig. \ref{match}) but NLO in
$k_+/M$ and $({\bf k}_{\perp}/M)^2$. We have checked that the coefficient of the logarithm coincides with that of the $(1-z)\log (1-z)$ term in the QCD
calculation \cite{earlier}, as it should.

The above means that the scale dependence of
the leading order contributions of the color-octet currents is of the same order as
the NNLO contributions in $\als$ of the color-singlet current, a calculation which is not available.
One might, alternatively, attempt to resum logs and use the NLO calculation \cite{kramer} as the
boundary condition. This log resummation is non-trivial. One must take into account the correlation of scales
inherent to the non-relativistic system \cite{LMR}, which in the framework of pNRQCD has been implemented in
\cite{pinedarg,currentrg}, and combine it with the resummation of Sudakov logs in the framework of SCET
\cite{luke,many,FL0,FL} (see also \cite{hautmann}).
Correlations within the various scales of SCET may start playing a role here
as well  \cite{MSS} . In any case, it should be clear that by only resumming Sudakov logs, as it has been done so far \cite{many}, one does
not resum all the logs arising in the color octet contributions of heavy quarkonium, at least in the weak coupling regime.

\label{scet}

\section{Application to the $\Upsilon (1S)$}\label{appl}

\begin{figure}
\centering
\includegraphics{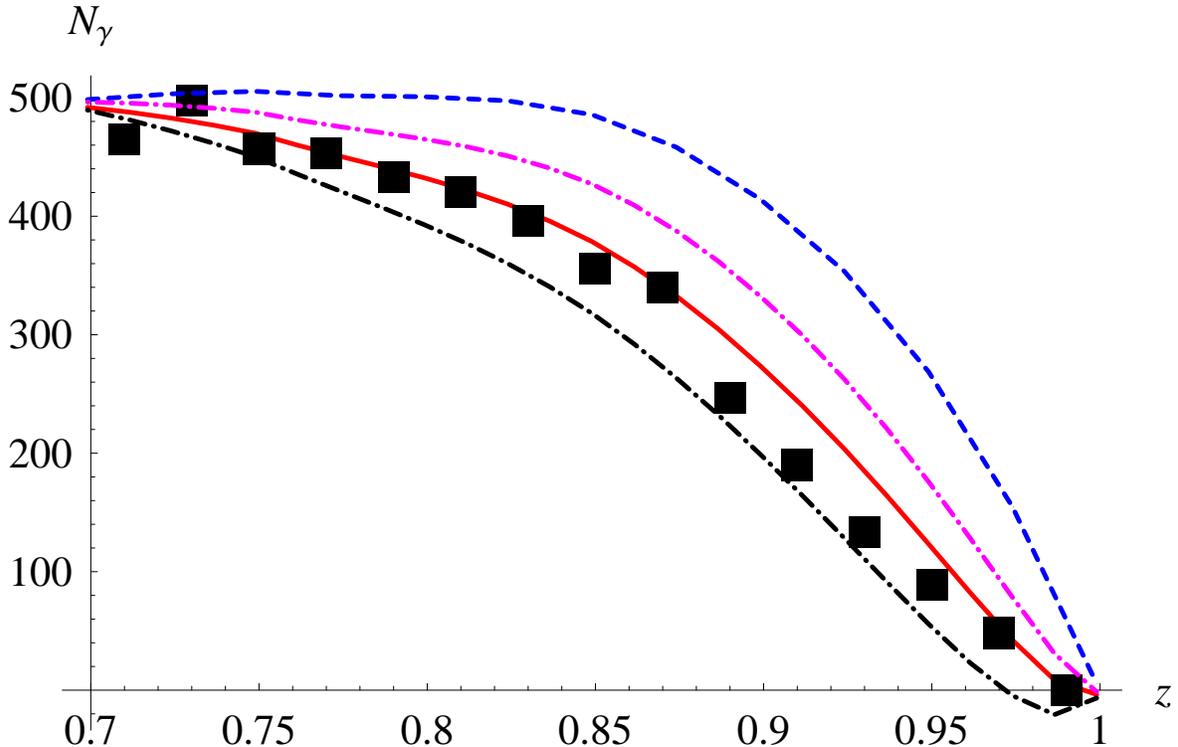}
\caption{\label{grafic}End point region of the photon spectrum in semi-inclusive $\Upsilon$ decay. The points are the CLEO data \cite{cleo},
the dashed line is the curve obtained in \cite{FL} and the solid and dot-dashed lines are our results. The solid line is obtained by setting $\mu=M\sqrt{(1-z)}$ (natural choice) and the dot-dashed lines are obtained
by setting $\mu=2M\sqrt{1-z}$ and $\mu=2^{-1}M\sqrt{1-z}$.}
\end{figure}

We apply here the results of section \ref{phespwcr} to the $\Upsilon (1S)$. There is good evidence that the  $\Upsilon (1S)$ state
can be understood as a weak coupling (positronium like) bound state \cite{yndurain,sumino,hamburg,pinedahamburg}. Hence, ignoring
$O\left(\Lambda_{\rm QCD}\right)$ in the shape functions, as we did in section \ref{phespwcr}, should be a reasonable approximation.
 In the analysis of \cite{FL} the effects of the octet shape functions were set to zero, so we expect to improve on their results.
 We plot in Fig. \ref{grafic} the CLEO data in the end point region \cite{cleo}, the curve obtained in \cite{FL} (dashed line) and
our curves (solid and dot-dashed lines). Our curves are obtained by adding to the results of \cite{FL}, which consist of the LL
 resummation of the singlet contributions only, our (MS) results for the color octet contributions (without LL resummation \cite{many}) and
setting the scale dependence to $\mu=M\sqrt{1-z}$ (solid line) and to $\mu=2^{\pm 1} M\sqrt{1-z}$ (dot-dashed lines). The first choice is the
most reasonable one according to the discussion in the previous section and the last ones are displayed in order to get the flavor
of the systematic errors.
We have used the following values for the masses and $\als$ in our plots: $m_b=4.8$ GeV, $M_{\Upsilon}=9.46$ GeV,
$\als\left(\mu_{h}\right)=0.216$, $\als\left(\mu_{s}\right)=0.32$ and $\als\left(\mu_{u}\right)=0.65$. $\mu_h \sim m$ stands for the hard scale and is to be used for the $\als$ arising from (3). $\mu_s \sim m\als$ stands for the soft scale and is to be used for the $\als$ participating in the bound state dynamics (in $E_1$, $I_S$, $I_P$ and $\psi_{10}({\bf 0})$). $\mu_u \sim m\als^2$ stands for the ultrasoft scale and is to be used for the $\als$ arising from the coupling of ultrasoft gluons.
It turns out that the color octet contribution is numerically enhanced and dominates over the color singlet one in the whole end-point region.
In order to compare with the experimental curve, the theoretical result must be convoluted with the experimental efficiency \cite{cleo} and the overall normalization must be taken as a free parameter.
By adjusting our curves to data around $z\sim 0.7$, we obtain an almost
perfect agreement in the whole end-point region ($z\in [0.7 , 1]$) for $\mu=M\sqrt{1-z}$ (solid line). A complete analysis, including systematic errors, is beyond the
scope of this paper. It would require either a NNLO matching or a NLO one \cite{kramer} with NLL resummation of the singlet current.
In addition one should estimate what the leading non-perturbative effects are.
In any case, it should be clear from our results that the introduction of a gluon mass \cite{field} is not necessary for the description
of the experimental data on the photon spectrum in the end-point region of $\Upsilon (1S)$ radiative decays.

\section{Discussion}\label{disc}

We would like to make a few remarks which stem from the details of our calculation. In the existing
 formulations of SCET, suitable scaling properties are assigned to the various modes. However the standard
assignments are violated in our case. The collinear gluons in Fig. \ref{dos} scale as $m(\lambda^2, 1, \lambda^2)$ rather
than $m(\lambda^2, 1, \lambda)$ ($\lambda=\sqrt{1-z}$), which is the assigned scaling in \cite{bauer,beneke,neubert}. This indicates that SCET should better be discussed in terms of UV (and IR) cut-offs for the relevant
modes, rather than scaling properties. This becomes particularly clear when we analyze the scaling of the
ultrasoft gluon in the same diagrams. If $m\als^2\sim M(1-z)$, it scales like $m(\lambda^2,\lambda^2,\lambda^2)$
($\lambda=\als$), which coincides with the standard scaling rules. If, however, $m\als^4\sim M(1-z)$, it scales
like $m(\lambda^4,\lambda^2,\lambda^{3})$, the typical scaling of the recently discovered
soft-collinear modes \cite{neubert}. Either scaling is properly described by the ultrasoft gluons of pNRQCD, since they are defined as the ones having
all four momenta much smaller than the soft scale $\sim m\als$ (UV cut-off), which is fulfilled in both cases. However, if one insisted to assign to the ultrasoft gluons a momentum scaling $m(\lambda^2,\lambda^2,\lambda^2)$ and not smaller (i.e. one is introducing an IR cut-off for them), then the situation $m\als^4\sim M(1-z)$ would require the introduction of new (ultra)soft-collinear modes scaling like $m(\lambda^4,\lambda^2,\lambda^{3})$ with an UV cut-off $\sim m\lambda^2$ for the ${+}$ and ${\perp}$ components. Whether it is convenient or not to make such splitting is a matter of debate \cite{neubert,new}. 

The case $m\als \sim M(1-z)$ has not been discussed. For the color-octet contributions, it requires a calculation in NRQCD, since if one attempts at
doing it from Fig. \ref{dos} one immediately realizes that the four momentum of the ultrasoft gluon is $\sim m\als$, a
region where pNRQCD is not applicable. Hence at NRQCD+SCET$_{\rm I}$ level (Section \ref{NS}) one should calculate a set of
diagrams involving a collinear and a soft gluon. The leading order contribution comes from the $S$-wave current
only and we have seen it to vanish.

We have refrained ourselves to put forward a Lagrangian for SCET which also holds for heavy quarkonium systems because
several issues, like the remarks made above, should be better understood. Clearly, as we have shown in this paper, one cannot simply take over the SCET for heavy-light systems and apply it to heavy quarkonium (for instance, one misses logs which
depend on the binding effects). Our analysis also indicates that it may be convenient to rephrase SCET in terms
of cut-offs rather than in terms of scaling properties of the various modes as it has been done so far. Then,
one would have to account properly for both the suitable cut-offs of pNRQCD and those of SCET.

\section{Conclusions}\label{concl}

Apart from making a few remarks, which we hope will be useful for an eventual construction of a SCET Lagrangian
adapted to heavy quarkonium systems, we have calculated the $S$- and $P$-wave octet shape functions in the
weak coupling regime. We have also discussed their scale dependence. The addition of these contributions to the
ones obtained in \cite{FL} makes the agreement with data for the end point photon spectrum of inclusive $\Upsilon (1S)$
decays almost perfect. As a by-product the NRQCD matrix elements
 $\left< \Upsilon (1S) \vert O_8 (^1 S_0) \vert \Upsilon (1S) \right>$ and
 $\left< \Upsilon (1S) \vert O_8 (^3 P_J) \vert \Upsilon (1S) \right>$ have also been calculated, in the weak
coupling regime, (see Appendix \ref{calcu}) and their scale dependence discussed.




\begin{acknowledgments}
We have benefited from informative discussions on SCET with C. Bauer, T. Becher, M. Beneke, S. Fleming, T. Mehen, M. Neubert and I. Stewart.
 J.S. thanks the Benasque Center of Physics program {\it Pushing the limits of QCD},
 and P. Bedaque for the opportunity to participate in the {\it Effective Summer in Berkeley},
where some of these discussions took place. Thanks are also given to D. Besson for providing the data of Ref. \cite{cleo}
and to Ll. Garrido for his help in handling it.
Finally, we thank A. Penin,  and, specially, A. Pineda for useful discussions on the Appendix A.
We acknowledge financial support from a CICYT-INFN 2003 collaboration contract,
the MCyT and Feder (Spain) grant FPA2001-3598, the
CIRIT (Catalonia) grant 2001SGR-00065 and the network EURIDICE (EU)
HPRN-CT2002-00311. X.G.T. acknowledges a FI fellowship from the Departament d'Universitats, Recerca i
Societat de la Informaci\'o of the Generalitat de Catalunya.
\end{acknowledgments}

\appendix

\section{Calculation of $\Upsilon (1S)$ NRQCD color octet matrix elements}\label{calcu}

\begin{figure}
\centering
\includegraphics{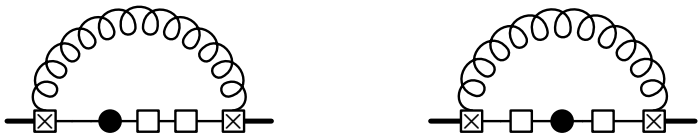}
\caption{\label{z2}Diagrams which require a ${\cal P}_1 (^3 S_1)$ operator for renormalization. The solid circle stands for either the
$O_8 (^1 S_0)$ or $O_8 (^3 P_J)$ operator, the crossed box for either the chromomagnetic (\ding{60} ) or chromoelectric
(\ding{182} ) interaction
in Fig. 1, the empty box for the octet Coulomb potential, and the thin solid lines for free $Q\bar Q$ propagators.}
\end{figure}
\begin{figure}
\centering
\includegraphics{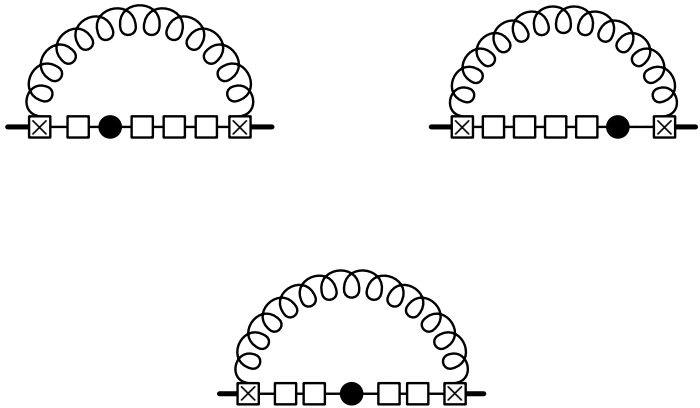}
\caption{\label{z4}Diagrams which require a $O_1 (^3 S_1)$ operator for renormalization. Symbols are as in Fig. \ref{z2}. }
\end{figure}

The calculation in Section \ref{cpns} can be easily taken over to provide a calculation of
$\left<\Upsilon(1S)\right\vert $ $\left. O_8(\phantom{}^1 S_0)\vert\Upsilon (1S) \right> $
and $\left<\Upsilon (1S)\vert O_8(\phantom{}^3 P_J)\vert\Upsilon (1S)\right>$
assuming that $m\als^2\gg \Lambda_{QCD}$ is a reasonable approximation for this system. Indeed, we only have to drop the delta function (which requires a further integration over $k_{+}$) and arrange for the suitable factors in (\ref{resonas}) and (\ref{ImTp}). We obtain
\begin{displaymath}
\left< \Upsilon (1S) \vert O_8 (^1 S_0) \vert \Upsilon (1S) \right> =-2T_F^2 (N_c^2-1)
\int_0^{\infty}dk_+S_{S}(k_+)
\end{displaymath}
\begin{equation}
\left< \Upsilon (1S) \vert O_8 (^3 P_J) \vert \Upsilon (1S) \right> =-{4(2J+1)T_F^2 (N_c^2-1)\over 3}
\int_0^{\infty}dk_+S_{P1}(k_+)
\end{equation}
where we have used
\be
\int_0^{\infty}dk_+S_{P2}(k_+)={2\over 3} \int_0^{\infty}dk_+S_{P1}(k_+)
\ee
The expressions above contain UV divergences which may be regulated in the same way as in Section \ref{cpns}, namely by
calculating the ultrasoft loop in $D$ dimensions. These divergences can be traced back to the diagrams in Fig. \ref{z2} and
Fig. \ref{z4}. Indeed, if we expand $I_{S}$ and $I_{P}$ for $z^{\prime}$ large, we obtain
\begin{displaymath}
I_S\sim m\sqrt{\frac{\gamma}{\pi}}\frac{\als N_c}{2}\left\{\frac{1}{{z^{\prime}}}+\frac{1}{{z^{\prime}}^2}(-1+2\lambda\ln2)+\frac{1}{{z^{\prime}}^3}\left(1-2\lambda+\frac{\lambda^2\pi^2}{6}\right)+\right.
\end{displaymath}
\begin{equation}\label{IStaylor}
\left.+\frac{1}{{z^{\prime}}^4}\left(-1+\lambda(2\ln2+1)+\lambda^2(-4\ln2)+\frac{3}{2}\zeta(3)\lambda^3\right) +{\cal O}(\frac{1}{{z^{\prime}}^5})\right\}
\end{equation}
\begin{displaymath}
I_P\sim \sqrt{\frac{\gamma^3}{\pi}}
{8\over 3}(2-\lambda)
\left\{\frac{1}{2{z^{\prime}}}+\left(-\frac{3}{4}+\lambda\left(-\frac{1}{4}+\ln2\right)\right)\frac{1}{{z^{\prime}}^2}+\right.
\end{displaymath}
\begin{displaymath}
\left.+\left(1-\lambda+\frac{1}{12}(-6+\pi^2)\lambda^2\right)\frac{1}{{z^{\prime}}^3}+\frac{1}{4}\left(-5+\lambda+\lambda^2(2-8\ln2)+8\lambda\ln2+\right.\right.
\end{displaymath}
\begin{equation}\label{IPtaylor}
\left.\left.+\lambda^3\left(-4\ln2+3\zeta(3)\right)\right)\frac{1}{{z^{\prime}}^4} +{\cal O}(\frac{1}{{z^{\prime}}^5})\right\}
\end{equation}
It is easy to see that only powers of $1/z^{\prime}$ up to order four may
give rise to divergences. Moreover, each power of $1/z^{\prime}$ corresponds to one Coulomb exchange.
Taking into account the result of the following integral,
\be
\int_0^\infty dk_+\int_0^\infty dx(2k_+x)^{-\varepsilon}\frac{1}{{z'}^{\alpha}}=2^{1-2\varepsilon}\left(\frac{\gamma^2}{m}\right)^{2-2\varepsilon}\frac{\Gamma^2(1-\varepsilon)}{\Gamma\left(\frac{\alpha}{2}\right)}\Gamma\left(\frac{\alpha}{2}+2\varepsilon-2\right)
\ee
we see that only the
$1/{z^{\prime}}^2$ and $1/{z^{\prime}}^4$ terms produce divergences.
The former correspond
to diagrams in Fig. \ref{z2} and the latter to Fig. \ref{z4}, which can be renormalized by the operators ${\cal P}_1 (^3 S_1)$ and
$O_1(^3S_1)$ respectively. It is again important to notice that these divergences are a combined effect of the ultrasoft
loop and quantum mechanics perturbation theory ({\it potential} loops \cite{benekesmirnov}) and hence it may not be clear at first sight if
they must be understood as ultrasoft (producing $\log \mu_u$ in the notation of Refs. \cite{pinedarg,currentrg}) or
potential (producing $\log \mu_p$ in the notation of Ref. \cite{pinedarg,currentrg}). In any case, the logarithms they
produce depend on the regularization and renormalization scheme used for both ultrasoft and potential loops.
Notice that the scheme we use in this work
 is not the standard one in pNRQCD \cite{lambpos,pinedarg,hamburg}. In the standard scheme the ultrasoft divergences
(anomalous dimensions) are identified by dimensionally regulating both ultrasoft and potential loops and subsequently taking
$D\rightarrow 4$ in the ultrasoft loop divergences only. If we did this in the present calculation we would obtain no ultrasoft
divergence. Hence, in the standard scheme there would be contributions to the potential anomalous dimensions only.
The singular pieces in our scheme are displayed below
\begin{displaymath}
\left.\left< \Upsilon (1S) \vert O_8 (^1 S_0) \vert \Upsilon (1S) \right>\right\vert_{\varepsilon\rightarrow 0} \simeq -{1\over \varepsilon}\left({2\gamma^2\over \mu m}\right)^{-2\varepsilon}{1\over 24} C_{F}^2 N_c \als (\mu_u)\left(C_f\als (\mu_s) \right)^4 {\gamma^3\over \pi^2}
\Bigg( 2+
\end{displaymath}
\begin{displaymath}
\left.
+\lambda \left[ -7-4\log 2\right]
+\lambda^2\left[ 4+8\log 2 + 4\log^2 2 + {\pi^2\over 3}\right]
 + \lambda^3\left[ -4\log^2 2-{\pi^2\over 3}-{3\over 2}\zeta (3)\right]\right)
\end{displaymath}
\begin{displaymath}
\left.\left< \Upsilon (1S) \vert O_8 (^3 P_J) \vert \Upsilon (1S) \right>\right\vert_{\varepsilon\rightarrow 0} \simeq -(2J+1){1\over \varepsilon}\left({2\gamma^2\over \mu m}\right)^{-2\varepsilon}{4\over 27} C_f\als (\mu_u)(C_f\als (\mu_s) )^2 {\gamma^5\over \pi^2} \times
\end{displaymath}
\begin{displaymath}
\times (2-\lambda )\left( -4 +
+\lambda\left[ {47\over 12}+5\log 2\right]+\lambda^2\left[ {5\over 6}-{2\pi^2\over 9}-{8\over 3}\log 2-{8\over 3}\log^2 2 \right]+\right.
\end{displaymath}
\begin{equation}
\left.+\lambda^3\left[ -{7\over 12}+{\pi^2\over 9}-{5\over 3}\log 2 + {4\over 3}\log^2 2 +{3\over 4}\zeta (3)\right] \right)
\end{equation}

\end{document}